# TRENDS IN FAST FEEDBACK R&D

Alessandro Drago

Istituto Nazionale di Fisica nucleare, Laboratori Nazionali di Frascati, Frascati, Italy.

*Abstract*

In this paper, starting from the basic description of the equation that governs the bunch motion and looking at the advances of the technology, three examples of feedback designs versus technology trend are presented and discussed. In particular the author compares three digital systems implemented or proposed for DAΦNE and other e+/e- accelerators. Descriptions of some relevant features are also done. Conclusions on the digital feedback design trend are reported.

## INTRODUCTION

In the last two decades, the impressive progress of the digital electronics has stunned and delighted designers and users. In the same years, feedback systems with increasing complexity have been designed to control the beam motion in lepton accelerators and colliders with more and more high currents.

In the following, some general consideration on fast bunch-by-bunch feedback systems are presented as well as the present digital electronics trends with special regards to commercially available DSP (Digital Signal Processor) and FPGA (Field Programmable Gate Array) technology. The synergy between development systems and beam-feedback simulators is also highlighted. Old and new DAΦNE bunch-by-bunch feedback systems are described with special attention to the last upgrade of the recent iGp system.

## BUNCH MOTION EQUATION

The n-th bunch can be described as an individual harmonic oscillator moving rigidly in the longitudinal plane (energy oscillations), or in the X, Y transverse panes (betatron oscillations) according to the classic harmonic equations. In the longitudinal plane, the dynamic of the system is described by the following equation [1]:

$$\ddot{\tau}_n + 2d_r \dot{\tau}_n + \omega_s^2 \tau_n = -\frac{\alpha_c e}{E_0 T_0} V_n^{wk}(t) \quad (1)$$

where $\tau_n$ is the arrival time (time delay) of the n-th bunch relative to the synchronous particle, $d_r$ is the natural radiation damping, $\omega_s$ is the natural (synchrotron) oscillation frequency, $\alpha_c$ is the momentum compaction, $E_0$ is the nominal energy, $eV_n^{wk}(t)/T_0$ is the rate of energy loss due to the superposition of the wake forces of the other bunches.

The action of the feedback consists in individual kicks to each bunch increasing the damping term $d_r$ by a voltage opposing to the wake field one. In presence of an active feedback system the equation (1) becomes:

$$\ddot{\tau}_n + 2d_r \dot{\tau}_n + \omega_s^2 \tau_n = \frac{\alpha_c e}{E_0 T_0} [V_n^{fb}(t) - V_n^{wk}(t)]$$

where $V_n^{fb}(t)$ represents the feedback kick applied to the bunch n-th at the time t.

## FEEDBACK MAIN BLOCKS

To introduce the discussion, in the following section we analyze, as example, the DAΦNE longitudinal feedback system (LFB) consisting of three main blocks (see fig.1); similar considerations can be done for transverse systems. The blocks are the following:

(i) A longitudinal pickup detecting a signal from the bunch passage and sending it to an analog front end followed by a programmable delay, to detect the error signal of each bunch. The function of the programmable delay is to synchronize the output signal of this block with the digital part.

(ii) A digital part, to manage separately the signal of every bunch with individual band-pass filters having a convenient gain and phase response. The global phase response of the feedback must give a 90 degrees phase shift at the dipole frequency to have a damping effect.

(iii) An analog back end (BE) followed by a second programmable delay, power amplifiers, and kicker. The BE programmable delay has the task of synchronizing the peak of the n-th kick with the passage of the n-th bunch through the kicker.

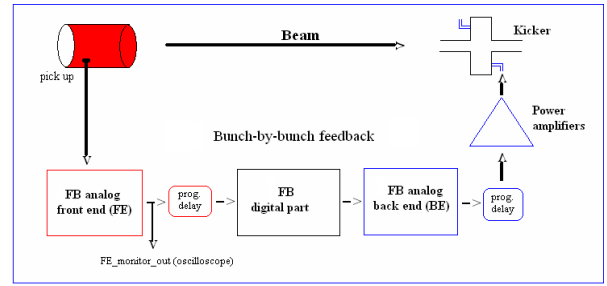

Fig.1 – DAΦNE longitudinal feedback main blocks

In the digital part, the FE output signal is sampled by an analog-to-digital converter working at RF frequency. After this task a demultiplexer separates the signal of each bunch in different slower computing paths.

A digital signal processor (DSP) farm is used to implement band-pass filters [finite impulse response (FIR) or infinite impulse response (IIR)]. Number of taps, gain with sign, center frequency, filter shape, and phase response are programmable by the user looking at the behaviour of the beam. The loaded filters are identical for all bunches, even if it is possible to run an "exception" filter for just one bunch.

In general, the IIR filter can be described by the following "conceptual" formula in which the output is function of the previous input and output bunch values:

$$y_n = \sum_{k=1}^{N} a_k y(n-k) + \sum_{j=0}^{M} b_j x(n-j)$$

Differently, implementing a FIR filter, the output is function only of the previous input bunch values:

$$y_n = \sum_{j=0}^{M} c_j x(n-j)$$

In general, there are infinite damping solutions for the coefficient $a_k$, $b_j$ and $c_j$ to be used to solve the motion equation see in the previous paragraph. The damping coefficients can be found experimentally or computed using a model. Amplitude and phase response versus frequency of a FIR filter implemented in the DSP farm for high current beam are shown in the fig. 2. The synchrotron frequency is drawn by a dotted line.

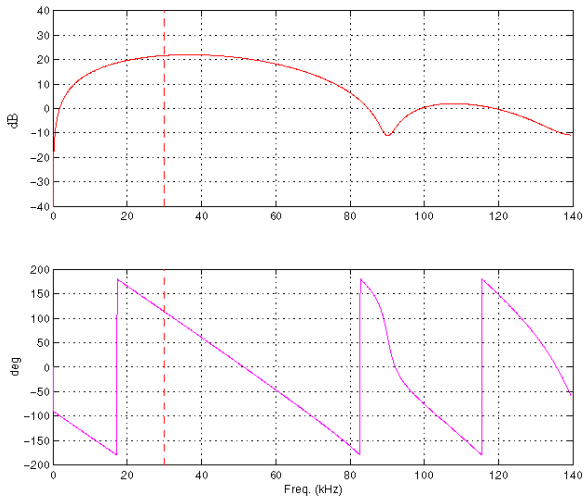

Fig.2 – FIR filter amplitude and phase response versus frequency

## PROGRESS IN DIGITAL PROCESSING

As well known, in the last two-three decades, impressive progresses have been done in the digital electronics. DSP (digital signal processor) units are single-chip microprocessors invented in the mid-80's to process mainly voice and audio signals (.1-10kHz bandwidth). DSP units have dedicated instruction set to perform real time (but at low frequency) signal processing.

As shown in the next paragraph, during the 90's a SLAC-LBNL-LNF collaboration has built a bunch by bunch longitudinal system (still used) designed around a 16 bits fixed point DSP by ATT.

In the second half of 90's, the designers watched a new powerful technological improvement: the FPGA (field programmable gate array) units were put on the market. As it is well known, the FPGA's are digital components with extremely high number of circuits in a single chip and the possibility to be reprogrammed "on the field" any time is necessary. The FPGA's became so powerful to begin to include PowerPc and many DSP units inside a single component.

The main FPGA producer, Xilinx Inc., foresees the impressing processor performance trend shown in the fig.3 [2]. The Xilinx Virtex-5 SX95T has 640 built-in 18x25 bits DSP allocated in only one chip running at 550Mhz. Powerful software environment makes possible to integrate beam-feedback simulators written in MATLAB language together with the FIR/IIR filter codes downloaded in the real hardware.

To compare the computing power now available, we should consider that the most recent version of digital feedback used at DAFNE is based on the Virtex-II that is shown on the low left side of the fig. 3.

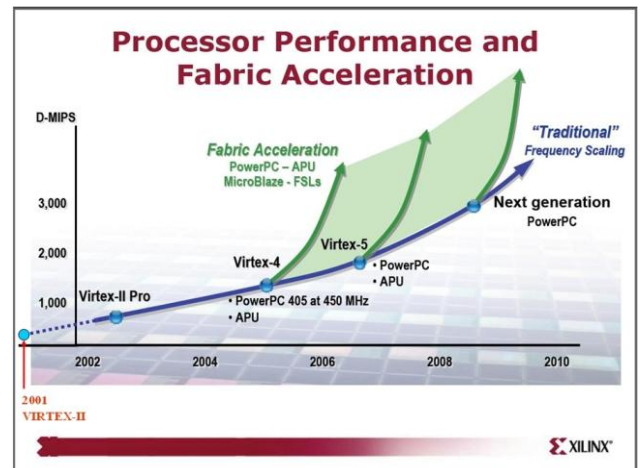

Fig.3 – Xilinx FPGA processor performance

## BUNCH-BY-BUNCH DIGITAL FEEDBACK VERSIONS

Looking back at the past feedback versions related to DAΦNE, we will find a first collaboration developing the longitudinal bunch-by-bunch digital feedback in the years 1992-1996 by a SLAC–LBNL-LNF team [3]. This system has still worked in the current year 2008 at PEP-II, DAFNE and ALS. It is based on a VXI crate and with up to 4 VME panels managing 5 DSP boards each. In every VME board 4 DSP units are allocated. In total the system can manage up to 80 DSP and each of them can elaborate up to 32 bunches. This feedback system was designed only to damp longitudinal oscillations and not transverse, because the betatron motion is generally present at too high frequencies.

In the September 2003 a new feedback design has been proposed for both longitudinal and transverse system by J.D.Fox and D.Teytelman at the ICFA'03 Workshop hold at Alghero [4]. This project, called GBoard, is based on at least four FPGA components, probably seven units in total, and is designed to be able to sample at 1.5 GS/sec using an 8 bits analog-to-digital converter. The processing speed is foreseen to manage the whole flux of data without need of downconversion for the input signals.

The "iGp" (integrated Gigasample processor) system has been developed by a collaboration of KEK-SLAC-LNF starting around 2002-2003 as a "small" prototype system for the "big" GBoard feedback system.

The iGp feedback system is a baseband bunch-by-bunch signal processing channel designed around a single Virtex-II FPGA by Xilinx.

It can be used for longitudinal and transverse feedback applications in storage rings as well as for bunch-by-bunch diagnostics. The "iGp" system has processed more than 5000 bunch signals sampled at ~500 MHz by an 8bits A/D converter. It uses EPICS as operator interface and distributed control system, working in Linux environment both in the IOC and in the client sides. MATLAB post-processing programs are used to analyze feedback performances and beam instabilities [5].

DAΦNE has 4 "iGp" feedback units currently running on the e+/e- transverse planes. At SLAC D.Teytelman and J.D.Fox have built ~15 units of the "iGp" system, some of them have been planned for working as bunch-by-bunch diagnostics systems. At KEK, M.Tobiyama and T.Obina have built a version the "iGp" system made in Japan, to be used in the longitudinal plane of the B-Factory and in the Photon Factory. Last but not least, at ALS (LBNL) one iGp unit is in phase of installation on the longitudinal plane replacing the old DSP-based system.

The iGp system is evolving fast and the last gateware (FPGA code) and software version has been tested at DAΦNE on the first week of April 2008 by D.Teytelman and myself. All the four iGp systems have been updated to the same software version based on a powerful Fedora rev. 8 Linux client personal computer.

A new important feature is a complete efficient control of the timing setup. This improvement has permitted to remove the Colby delay lines in the front end and the old analog delay lines in the back end stage.

The front-end amplifiers have been also removed in all the transverse planes to have a smaller crosstalk between bunches by using all the frequency band of the sharp input beam pulse. Now the pickup signal, after passing through the H9 hybrids making the difference operation, enters directly in the iGp unit.

The iGp feedback has also, as further features, many diagnostic tools included in the system. As example from the Epics panels it possible to make:
- input signal record with sw trigger from operator i/f;
- grow/damp record with sw trigger from operator i/f;
- data record during the injection using a hw trigger coming from the DAΦNE timing system;

From the off-line MATLAB environment it is possible:
- to store recorded data in a time-stamped database creating two files in .mat format;
- to make post-processing analysis on the database;
- to do beam modal analysis (mode # and grow rate);
- to make injection transient data analysis to study the injection kicker effects versus kicker setup and timing;
- to do bunch-by-bunch tune spread analysis.

As example, in the horizontal e+ plane, grow/damp data have been recorded at 355 mA; the off-line data analysis has shown the presence of a strong -1 mode and an extremely fast damping of the feedback of the order of 2.5 microseconds rate.

The bunch-by-bunch tune analysis, made off-line on data recorded, has shown that a large tune spread along the train can be observed in the horizontal plane while a much smaller tune spread is present in the vertical plane.

Another interesting real time feature is shown in the fig.4. The frequency power spectrum of the beam signal (the low square on the right) is able to show the rejecting frequency corresponding to the betatron tune.

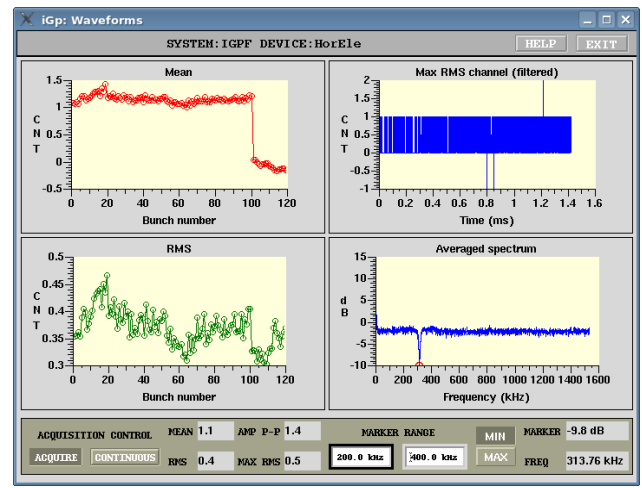

Fig.4 – iGp Waveform panel

## CONCLUSION

The FPGA (Field Programmable Gate Array) technology upgrades very fast its performances: this trend should be considered in designing future feedback projects to foresee new features and more powerful capabilities. The iGp (integrated Gigasample processor) feedback is a single FPGA system with many new interesting features. Powerful diagnostics inside the system can help to understand beam current limits and evaluate feedback performances. In DAFNE e+ horizontal plane an extremely fast mode -1 and a large tune spread limit the beam stability and the storable beam current.